\documentclass[aps,preprint,amsmath,amssymb]{revtex4}
\usepackage{graphicx}
\begin{document}

\title{Study of anomalous $WW\gamma\gamma$ coupling sensitivity at the Compact Linear Collider}

\author{M. Koksal}
\email[]{mkoksal@cumhuriyet.edu.tr} \affiliation{Department of
Physics, Cumhuriyet University, 58140, Sivas, Turkey}

\begin{abstract}
The Compact Linear Collider (CLIC) is one of the most popular linear colliders, planned to realize $e^{-}e^{+}$ collisions in three energy stages of $0.5$, $1.5$, and $3$ TeV. It has an energy scale never reached by any existing lepton collider. In this study,
we present the sensitivity studies of the $WW\gamma\gamma$ anomalous quartic gauge boson
coupling (aQGC) in the three different processes $e^{+}e^{-}\rightarrow W^{-} W^{+}\gamma$,
$e^{+}e^{-} \rightarrow e^{+}\gamma^{*} e^{-} \rightarrow e^{+} W^{-} \gamma \nu_{e}$, and $e^{+}e^{-} \rightarrow e^{+}\gamma^{*} \gamma^{*} e^{-} \rightarrow e^{+} W^{-} W^{+} e^{-}$  at the CLIC. We obtained $95\%$ confidence level limits on the aQGC parameters $\frac{a_{0}}{\Lambda^{2}}$ and $\frac{a_{c}}{\Lambda^{2}}$ at various integrated luminosities and center-of-mass energies. The best limits obtained on the aQGCs $\frac{a_{0}}{\Lambda^{2}}$ and $\frac{a_{c}}{\Lambda^{2}}$ in these processes are at the order of $10^{-2}$ TeV$^{-2}$ and $10^{-1}$ TeV$^{-2}$, respectively.
\end{abstract}

\maketitle

\section{Introduction}

In the Standard Model (SM) of particle physics, triple and quartic interactions of the gauge bosons have not been determined with
great accuracy. The possible deviation of gauge boson self-interactions from the SM predictions will be an important sign of new physics beyond the SM.
Assuming that the energy scale associated with the new physics is sufficiently high compared to the masses of the gauge bosons, we are motivated to use Effective Lagrangian method \cite{effek}, which is model independent and can converge to the SM at lower energy scale. The aQGCs are parameterized by effective operators which do not induce new trilinear gauge couplings. Therefore, we consider that aQGCs can be examined independently from the effects induced by trilinear gauge couplings.

In the literature, the aQGC operators have been described by either linear or non-linear effective Lagrangians. While non-linear effective Lagrangians have dimension 6, linear effective Lagrangians have dimension 8. However, the dimension 6 Lagrangian only contains vertices with two photons. Assuming that there is no Higgs boson in the low energy spectrum we employ a nonlinear representation of the spontaneously broken $SU(2)\times U(1)$ gauge symmetry. C and P conserving non-linear effective Lagrangian for $WW\gamma\gamma$ aQGCs are given by
\cite{lag2,lag3}

\begin{eqnarray}
L=L_{0}+L_{c},
\end{eqnarray}

\begin{eqnarray}
L_{0}=\frac{-\pi\alpha}{4}\frac{a_{0}}{\Lambda^{2}}F_{\mu \nu}F^{\mu \nu} W_{ \alpha}^{(i)} W^{(i) \alpha},
\end{eqnarray}

\begin{eqnarray}
L_{c}=\frac{-\pi\alpha}{4}\frac{a_{c}}{\Lambda^{2}}F_{\mu \alpha}F^{\mu \beta} W^{(i)\alpha} W_{\beta}^{(i)}
\end{eqnarray}
where $W^{(i)}$ is $SU(2)_{W}$ triplet, $F_{\mu \nu}=\partial_{\mu}A_{\nu}-\partial_{\nu}A_{\mu}$ is the tensor for electromagnetic field strength, and $a_{0}$ and $a_{c}$ are the dimensionless. $\Lambda$ is a mass-dimension parameter associated with the energy scale of the new degrees of freedom which have been integrated out.

In addition, the dimension 8 Lagrangian induces quartic couplings among the neutral gauge bosons and it is
obtained by using a linear representation of $SU(2)\times U(1)$ gauge symmetry that is broken by the conventional SM Higgs mechanism. However,
after the recent discovery of a new particle which is consistent with the SM
Higgs boson, it becomes important to examine aQGCs based on linear effective
Lagrangian. On the other hand, since many aQGC-related
studies have been generally investigated using
non-linear effective Lagrangian, which we also choose in our aQGC study. In addition, our limits can compare with the results of the linear parameterizations which is published by CMS collaboration \cite{ilk}. For this reason, the non-linear effective Lagrangian can be expressed by the linear effective Lagrangian \cite{ww1,ww2,ww3}. Therefore, aQGC parameters obtained from non-linear operators can be easily translated into parameters of linear operator.

In effective Lagrangian approach, the cross
sections of aQGC processes increase with the coupling strength, and hence unitarity is violated for sufficiently high energy collisions. As stated by Ref. \cite{cin}, a plausible choice of form factor can really provide the unitarity requirement. Form factor formalism is commonly described by the formula \cite{ebol,D0}

\begin{eqnarray}
a_{0,c}\rightarrow\frac{a_{0,c}}{(1+\hat{s}/\Lambda^{2})^{n}}
\end{eqnarray}
where $\Lambda$ represents new physics scale. However, because of selection of form factor is arbitrary in the literature, we will examine aQGC parameters without using any form factor in this study.

In the presence of the effective Lagrangian in Eqs.(2-3), the vertex functions for $W^{+} (k_{1}^{\mu}) W^{-}(k_{2}^{\nu}) \gamma (p_{1}^{\alpha}) \gamma (p_{2}^{\beta})$ are obtained respectively by

\begin{eqnarray}
i\frac{2\pi\alpha}{\Lambda^{2}}a_{0}g_{\mu\nu}[g_{\alpha\beta}(p_{1}.p_{2})-p_{2\alpha}p_{1\beta}],
\end{eqnarray}

\begin{eqnarray}
&&i\frac{\pi\alpha}{2\Lambda^{2}}a_{c}[(p_{1}.p_{2})(g_{\mu\alpha}g_{\nu\beta}+g_{\mu\beta}g_{\alpha\nu})+g_{\alpha\beta}(p_{1\mu}p_{2\nu}+p_{2\mu}p_{1\nu}) \nonumber \\
&&-p_{1\beta}(g_{\alpha\mu}p_{2\nu}+g_{\alpha\nu}p_{2\mu})-p_{2\alpha}(g_{\beta\mu}p_{1\nu}+g_{\beta\nu}p_{1\mu})].
\end{eqnarray}

$WW\gamma\gamma$ aQGCs were examined in OPAL, D0 and CMS experiments. The $95\%$ C. L. limits on aQGCs $\frac{a_{0}}{\Lambda^{2}}$ and $\frac{a_{c}}{\Lambda^{2}}$ through the processes $e^{+}e^{-}\rightarrow W^{-} W^{+}\gamma$ and $e^{+}e^{-}\rightarrow \nu \bar{\nu} \gamma \gamma$ by the OPAL collaboration at LEP2 collider are given by $-2.0\, \times 10^{4} \textmd{TeV}^{-2}<\frac{a_{0}}{\Lambda^{2}}<2.0 \times 10^{4}$ TeV$^{-2}$, $-5.2\, \times 10^{4} \textmd{TeV}^{-2}<\frac{a_{c}}{\Lambda^{2}}<3.7 \times 10^{4}$ TeV$^{-2}$ \cite{OPAL}. Also, the current limits on aQGCs $\frac{a_{0}}{\Lambda^{2}}$ and $\frac{a_{c}}{\Lambda^{2}}$ have been provided through the processes $p\, p\, (\bar{p})\rightarrow p\,  \gamma^{*} \gamma^{*} p \, (\bar{p})\rightarrow p W^{-} W^{+} p \, (\bar{p})$ by D0 collaboration at the Tevatron \cite{D1} and CMS collaboration at the Large Hadron Collider (LHC) \cite{CMS}. The aQGC limits obtained by D0 collaboration are
\begin{eqnarray}
 -4.3\,\times 10^{2} \textmd{TeV}^{-2}<\frac{a_{0}}{\Lambda^{2}}<4.3 \times 10^{2} \, \textmd{TeV}^{-2},
\end{eqnarray}
\begin{eqnarray}
 -1.6\, \times 10^{3} \textmd{TeV}^{-2}<\frac{a_{c}}{\Lambda^{2}}<1.5 \times 10^{3} \, \textmd{TeV}^{-2}.
\end{eqnarray}

The most restrictive limits on the parameters of $WW\gamma\gamma$ aQGCs were determined by CMS collaboration
\begin{eqnarray}
-4\, \textmd{TeV}^{-2}<\frac{a_{0}}{\Lambda^{2}}<4 \,\textmd{TeV}^{-2},
\end{eqnarray}
\begin{eqnarray}
-15\, \textmd{TeV}^{-2}<\frac{a_{c}}{\Lambda^{2}}<15 \,\textmd{TeV}^{-2}.
\end{eqnarray}

In the literature, $WW\gamma\gamma$ aQGCs at the linear colliders were studied through the processes $e^{+}e^{-}\rightarrow VVV$ \cite{ee1,ee2,stir,ee3}, $e^{+}e^{-}\rightarrow VVFF$ \cite{stt}, $e\gamma\rightarrow VVF$ \cite{sato,sato1}, and $\gamma\gamma\rightarrow VVV$ \cite{mem,mem1}, where $V=Z,W^{\pm}$ or $\gamma$ and $F = e$ or $\nu$. In addition, these couplings were examined at the LHC via the processes $pp\rightarrow p \gamma^{*} \gamma^{*} p \rightarrow p W^{-} W^{+} p$ \cite{had,had1}, $pp\rightarrow p \gamma^{*} p \rightarrow p W \gamma q X$ \cite{had2}, $pp\rightarrow W\gamma\gamma$ \cite{mur,ebol,cin} and $qq\rightarrow qq \gamma \gamma$ \cite{had3}.

In this study, we probe the $WW\gamma\gamma$ aQGCs
by analyzing three different processes $e^{+}e^{-}\rightarrow W^{-} W^{+}\gamma$,
$e^{+}e^{-} \rightarrow e^{+}\gamma^{*} e^{-} \rightarrow e^{+} W^{-} \gamma \nu_{e}$, and $e^{+}e^{-} \rightarrow e^{+}\gamma^{*} \gamma^{*} e^{-} \rightarrow e^{+} W^{-} W^{+} e^{-}$ at the CLIC.

The LHC may not be the best platform to investigate genuine quartic gauge couplings due to remnants of usual $pp$ deep inelastic processes. On the other hand, since $e^{-}$ and $e^{+}$ are fundamental particles, lepton colliders can determine aQGC parameters with much higher precision measurements than hadron colliders. The CLIC, which has high energy and luminosity, is one of the foreseen linear colliders. It is designed to be running at three center-of-mass energy stages: $0.5$, $1.5$, and $3$ TeV \cite{17}.  The fundamental parameters of the three energy stages are given in Table I. In the CLIC's first energy stage, it provides good opportunities for the precise measurements of different observables and characteristics of the SM Higgs boson, top quark and gauge bosons, etc.. The second phase will further open the window to the discovery of new physics beyond the SM. In addition, Higgs boson properties such as the Higgs self-coupling and rare Higgs decay modes will be examined. The third phase, which has a maximum energy of $3$ TeV, is considered to be able to make the most precise measurements of the SM, and to directly determine the pair-production of new heavy particles with masses up to $1.5$ TeV \cite{18}.

The linear $e^{-}e^{+}$ collider also has $e\gamma$ and $\gamma\gamma$ run modes to examine the new physics search. In $e\gamma$ and $\gamma\gamma$ run modes, high energy real photons can be obtained by converting the incoming leptons beam into a photon
beam via the Compton backscattering mechanism \cite{las1,las2,las3,yeni,yeni1}.
In addition, the linear collider allows us to study photon-induced $e\gamma^{*}$ and $\gamma^{*} \gamma^{*}$ reactions arising from almost real photons. Here, $\gamma^{*}$ is emitted by any of the incoming lepton immediately after it collides with the other lepton. Hence, it is possible to investigate $e\gamma^{*}$ and $\gamma^{*} \gamma^{*}$ collisions at the CLIC. The photons in these processes have been described as a suitable framework by the Weizsaecker-Williams approximation (WWA) \cite{WWA,WWA1,WWA2,WWA3,WWA4,WWA5}. WWA is used to define processes of electro-production in the case of very small angle of $e^{-}$ or $e^{-}$ scattering. In this instance, since the virtuality of photon emitted from incoming leptons in WWA is very low, it is assumed to be on the mass shell. It gives a possibility to reduce the process of electro-production to the photo-production one with an appropriate photon spectrum.
As a result, $e\gamma^{*}$ and $\gamma^{*} \gamma^{*}$ processes are produced in a natural way from the $e^{-}e^{+}$ process itself. The diagrams defining these processes are presented in Fig. $1$ and $2$. In the literature, photon-induced reactions through the WWA have been widely examined at the LEP, Tevatron, and LHC \cite{399,400,401,402,43,44,45,q1,q2,q3,q4,q5,q6,q7,kok,kok2,4777,46,47,kok22,kok1,q8,q9,q10,q11,q12,q13}.

\section{Cross Sections}

The processes $e^{+}e^{-}\rightarrow W^{-} W^{+}\gamma$ and the subprocesses $e^{-}\gamma^{*} \rightarrow W^{-} \gamma \nu_{e}$, and $\gamma^{*} \gamma^{*}\rightarrow W^{-} W^{+}$ at the CLIC are described by tree-level Feynman diagrams in Figs. $3$-$5$. We can see
from these figures that only one of these diagrams involves anomalous $WW\gamma\gamma$ vertex while the others show the contributions arising from the SM. We have used the COMPHEP-4.5.1 program for numerical calculations in this study \cite{comp}. In our study, only one aQGC coefficient is set to 0 at one time while the other one is set to non-zero. The total cross sections as functions of $\frac{a_{0}}{\Lambda^{2}}$ and $\frac{a_{c}}{\Lambda^{2}}$ aQGCs for these processes at the CLIC with $\sqrt{s}=0.5,1.5$ and $3$ TeV are given in Figs. $6$-$14$.  We observe from Figs. $6$-$14$ that the cross section values induced by non-zero $\frac{a_{0}}{\Lambda^{2}}$ are larger than the values of the $\frac{a_{c}}{\Lambda^{2}}$ coupling. Hence, the obtained limits on the $\frac{a_{0}}{\Lambda^{2}}$ coupling are anticipated to be
more restrictive than the limits on $\frac{a_{c}}{\Lambda^{2}}$ coupling.

\section{Limits on the aQGCs}

During statistical analysis, we determined $95\%$ C.L. limits on the aQGC parameters $\frac{a_{0}}{\Lambda^{2}}$ and $\frac{a_{c}}{\Lambda^{2}}$
using a simple one-parameter $\chi^{2}$ analysis. The $\chi^{2}$ analysis is defined by the following formula

\begin{eqnarray}
\chi^{2}=\left(\frac{\sigma_{SM}-\sigma_{NP}}{\sigma_{SM}\delta_{stat}}\right)^{2}
\end{eqnarray}
where $\sigma_{SM}$ is the SM cross section, $\sigma_{NP}$ is the total cross section with aQGC, $\delta_{stat}=\frac{1}{\sqrt{N}}$ is the statistical
error: $N$ is the number of events. First, the number of events for the process $e^{+}e^{-}\rightarrow W^{-} W^{+}\gamma$
is given by

\begin{eqnarray}
N= L_{int} \times \sigma_{SM} \times BR(W\rightarrow q\bar{q}')\times BR(W\rightarrow \ell \nu_{\ell})
\end{eqnarray}
where $L_{int}$ is the integrated luminosity and $\sigma_{SM}$ is the SM cross section.
The $W$ boson is heavy enough to decay both hadronically and leptonically. It decays approximately
$21.5\%$ of the time leptonically (for electron or muon) and $67.6\%$ of the time to hadrons \cite{WB}. So we consider
one of the $W$ bosons decays leptonically and the other hadronically for the signal. Therefore,
we consider that the branching ratio of the $W$ bosons pairs in the final state to be $BR = 0.145$. In addition, we impose the acceptance cuts on the pseudorapidity  $|\eta^{\,\gamma}|<2.5$ and the transverse momentum $p_{T}^{\,\gamma}>20 \:$ GeV  for the final state photon. After applying these cuts, the SM background cross sections in the process $e^{+}e^{-}\rightarrow W^{-} W^{+}\gamma$, we obtain as $0.165$ pb at $\sqrt{s}=0.5$ TeV, $0.06$ pb at $\sqrt{s}=1.5$ TeV, and $0.026$ pb at $\sqrt{s}=3$ TeV.

In Table II, we calculate limits on the aQGC parameters $\frac{a_{0}}{\Lambda^{2}}$ and $\frac{a_{c}}{\Lambda^{2}}$ for various integrated luminosities and center-of-mass energies of the process $e^{+}e^{-}\rightarrow W^{-} W^{+}\gamma$. We observe from Table II that the obtained limits on the aQGCs via the process $e^{+}e^{-}\rightarrow W^{-} W^{+}\gamma$ with $\sqrt{s}=0.5$ TeV are less restrictive than the best limits attained by the current experimental limits. On the other hand, our limits at $1.5$ and $3$ TeV of center-of-mass energies of the same process are, even with low integrated luminosities, more restrictive than the experimental limits. Especially, we have found the limits of the $\frac{a_{0}}{\Lambda^{2}}$ aQGC as
$[-0.10;\, 0.07]$ TeV$^{-2}$ while the limits on $\frac{a_{c}}{\Lambda^{2}}$ aQGC as $[-0.18;\, 0.13]$ TeV$^{-2}$ at the integrated luminosity of $590$ fb$^{-1}$ and center-of-mass energy of $3$ TeV.

In the second analysis, the number of events for the process $e^{+}e^{-} \rightarrow e^{+}\gamma^{*} e^{-} \rightarrow e^{+} W^{-} \gamma \nu_{e}$
is obtained as

\begin{eqnarray}
N= L_{int} \times \sigma_{SM} \times BR(W\rightarrow q\bar{q}').
\end{eqnarray}
Here, the $W$ boson can decay leptonically, but in this case, it will introduce a great uncertainty due to the production of two neutrinos in the final state of our process.
For this reason, we consider the hadronic decay of the $W$ boson. Also, we apply the cuts $p_{T}^{\,\gamma}>20 \:$ GeV and $|\eta^{\,\gamma}|<2.5$ for the photon in the final state. Therefore, we calculate the SM cross sections as $0.021$ pb at $\sqrt{s}=0.5$ TeV, $0.124$ pb at $\sqrt{s}=1.5$ TeV, and $0.227$ pb at $\sqrt{s}=3$ TeV.
We show the limits on the $\frac{a_{0}}{\Lambda^{2}}$ and $\frac{a_{c}}{\Lambda^{2}}$ aQGCs for the process $e^{+}e^{-} \rightarrow e^{+}\gamma^{*} e^{-} \rightarrow e^{+} W^{-} \gamma \nu_{e}$ at various integrated luminosities and center-of-mass energies in Table III. As we can see from this table, the best limits obtained on $\frac{a_{0}}{\Lambda^{2}}$ and $\frac{a_{c}}{\Lambda^{2}}$ aQGCs through process $e^{+}e^{-} \rightarrow e^{+}\gamma^{*} e^{-} \rightarrow e^{+} W^{-} \gamma \nu_{e}$ are found to be $[-0.41;\, 0.40]$ TeV$^{-2}$ and $[-0.48;\, 0.68]$ TeV$^{-2}$, respectively for $L_{int}=590$ fb$^{-1}$ and $\sqrt{s}=3$ TeV at the CLIC.

Finally, for the process $e^{+}e^{-} \rightarrow e^{+}\gamma^{*} \gamma^{*} e^{-} \rightarrow e^{+} W^{-} W^{+} e^{-}$, in order to get promising results the number of events is obtained as

\begin{eqnarray}
N= L_{int} \times \sigma_{SM} \times BR(W\rightarrow q\bar{q}')\times BR(W\rightarrow \ell \nu_{\ell}).
\end{eqnarray}
Here, we find the SM cross sections to be $0.202$ pb for $\sqrt{s}=0.5$ TeV, $1.458$ pb for $\sqrt{s}=1.5$ TeV, and $3.199$ pb for $\sqrt{s}=3$ TeV.
In Table IV, we give the limits of the $\frac{a_{0}}{\Lambda^{2}}$ and $\frac{a_{0}}{\Lambda^{2}}$ aQGCs for $e^{+}e^{-} \rightarrow e^{+}\gamma^{*} \gamma^{*} e^{-} \rightarrow e^{+} W^{-} W^{+} e^{-}$ at CLIC at $\sqrt{s}=0.5, 1.5$ and $3$ TeV.
As shown in all these tables, we realize that the sensitivities of $\frac{a_{0}}{\Lambda^{2}}$ and $\frac{a_{c}}{\Lambda^{2}}$ aQGCs are rapidly enhanced  when the center-of-mass energy of the CLIC increases.

In addition, the combined limits obtained on $\frac{a_{0}}{\Lambda^{2}}$ and $\frac{a_{c}}{\Lambda^{2}}$ aQGCs through three different processes with the integrated luminosity of $L_{int}=590$ fb$^{-1}$ at the CLIC with $\sqrt{s}=3$ TeV are calculated as $[-0.21;\, 0.19]$ TeV$^{-2}$ and $[-0.39; \, 0.38]$ TeV$^{-2}$, respectively. The combined limits are worse than the limits derived in the process $e^{+}e^{-}\rightarrow W^{-} W^{+}\gamma$ since the other two processes have worse limits.

\section{Conclusions}

The CLIC is a future linear collider at the planning stage. Having it running with high energy and high luminosity is quite important new physics research beyond the SM. $WW\gamma\gamma$ aQGCs are defined by dimension 6 effective Lagrangians, which have very strong energy dependences. Hence, $WW\gamma\gamma$ cross section with anomalous couplings interactions has a higher momentum dependence than the SM cross section. We can easily understand that the contribution to the cross section of aQGCs quickly increases with center-of-mass energy. The analyzed processes that have very high energy and clean experimental environment such as CLIC, are anticipated to be more sensitive to aQGCs. Therefore, we are motivated to study the the aQGC parameters $\frac{a_{0}}{\Lambda^{2}}$ and $\frac{a_{c}}{\Lambda^{2}}$ through the processes $e^{+}e^{-}\rightarrow W^{-} W^{+}\gamma$,
$e^{+}e^{-} \rightarrow e^{+}\gamma^{*} e^{-} \rightarrow e^{+} W^{-} \gamma \nu_{e}$, and $e^{+}e^{-} \rightarrow e^{+}\gamma^{*} \gamma^{*} e^{-} \rightarrow e^{+} W^{-} W^{+} e^{-}$ at the CLIC. We have shown that the best limits obtained on aQGCs $\frac{a_{0}}{\Lambda^{2}}$ and $\frac{a_{c}}{\Lambda^{2}}$ in these processes are at the order of $10^{-2}$ and $10^{-1}$ TeV$^{-2}$, respectively. Since our limits are derived based on the total cross section with only statistical uncertainty, they is significantly better than the limits derived in experimental studies.

\pagebreak

\pagebreak

\begin{figure}
\includegraphics[width=0.53\columnwidth]{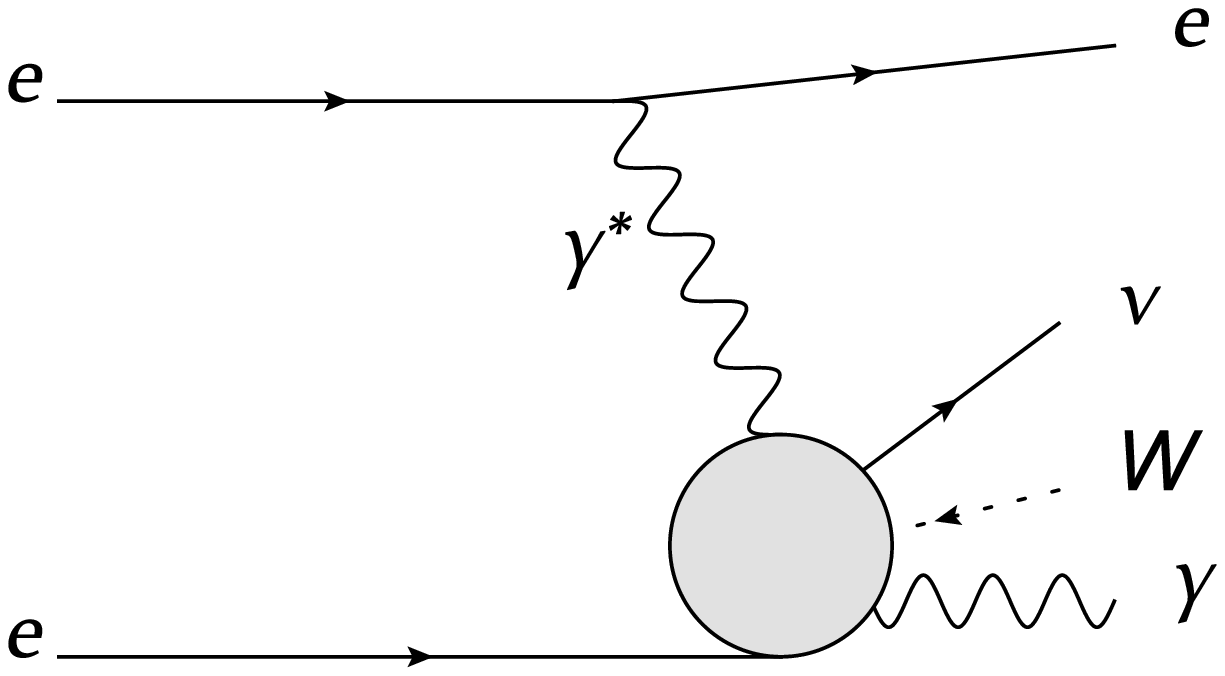}
\caption{The diagram defining the process $e^{+}e^{-}\rightarrow e^{+}\gamma^{*} e^{-}\rightarrow e^{+}W^{-} \gamma \nu_{e}$.
\label{fig1}}
\end{figure}

\begin{figure}
\includegraphics [width=0.5\columnwidth] {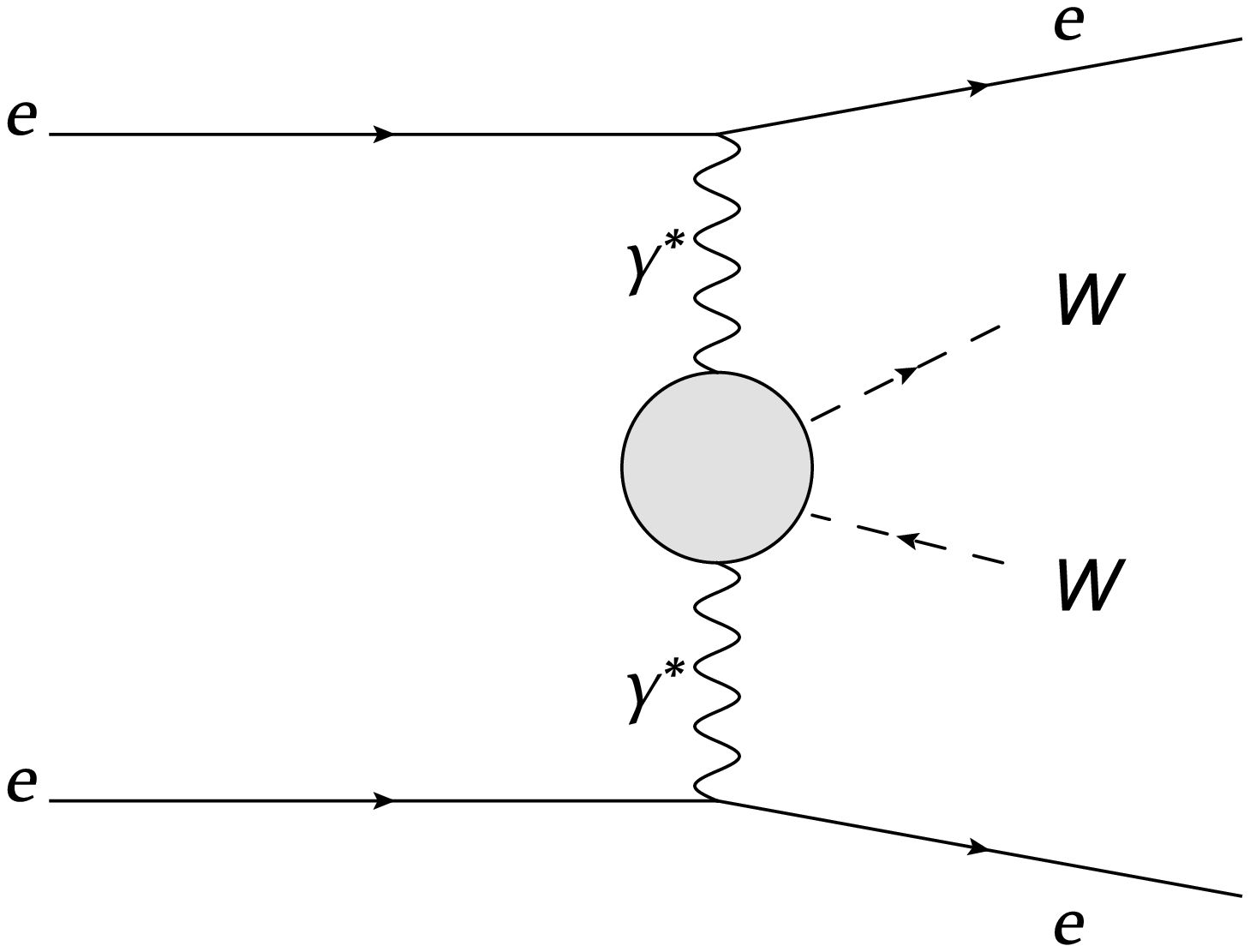}
\caption{The diagram defining the process $e^{+}e^{-} \rightarrow e^{+}\gamma^{*} \gamma^{*} e^{-} \rightarrow e^{+}W^{+} W^{-}e^{-}$.
\label{fig2}}
\end{figure}

\begin{figure}
\includegraphics{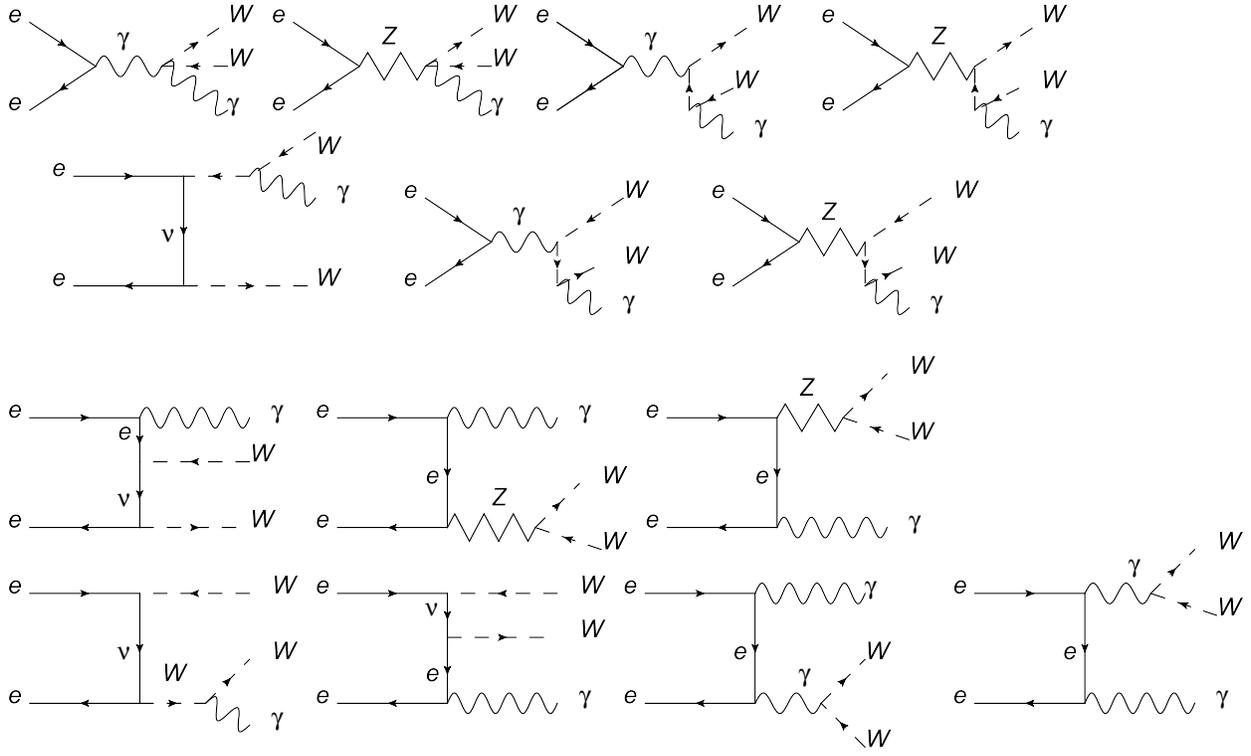}
\caption{Tree-level Feynman diagrams for the process $e^{+}e^{-}\rightarrow W^{-} W^{+}\gamma$.
\label{fig3}}
\end{figure}

\begin{figure}
\includegraphics{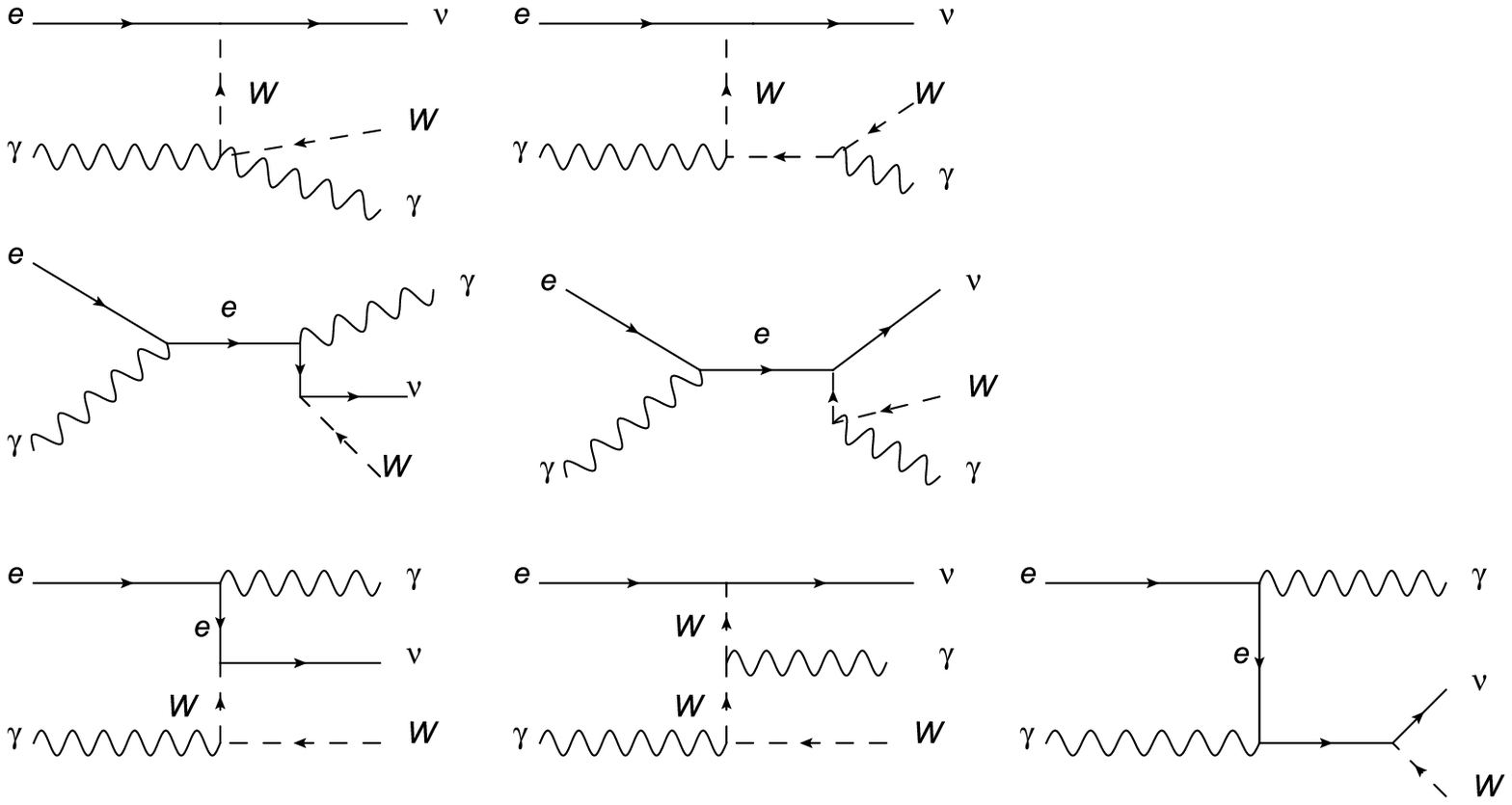}
\caption{Tree-level Feynman diagrams for the subprocess $e^{-}\gamma^{*} \rightarrow W^{-} \gamma \nu_{e}$.
\label{fig4}}
\end{figure}

\begin{figure}
\includegraphics [width=0.7\columnwidth]{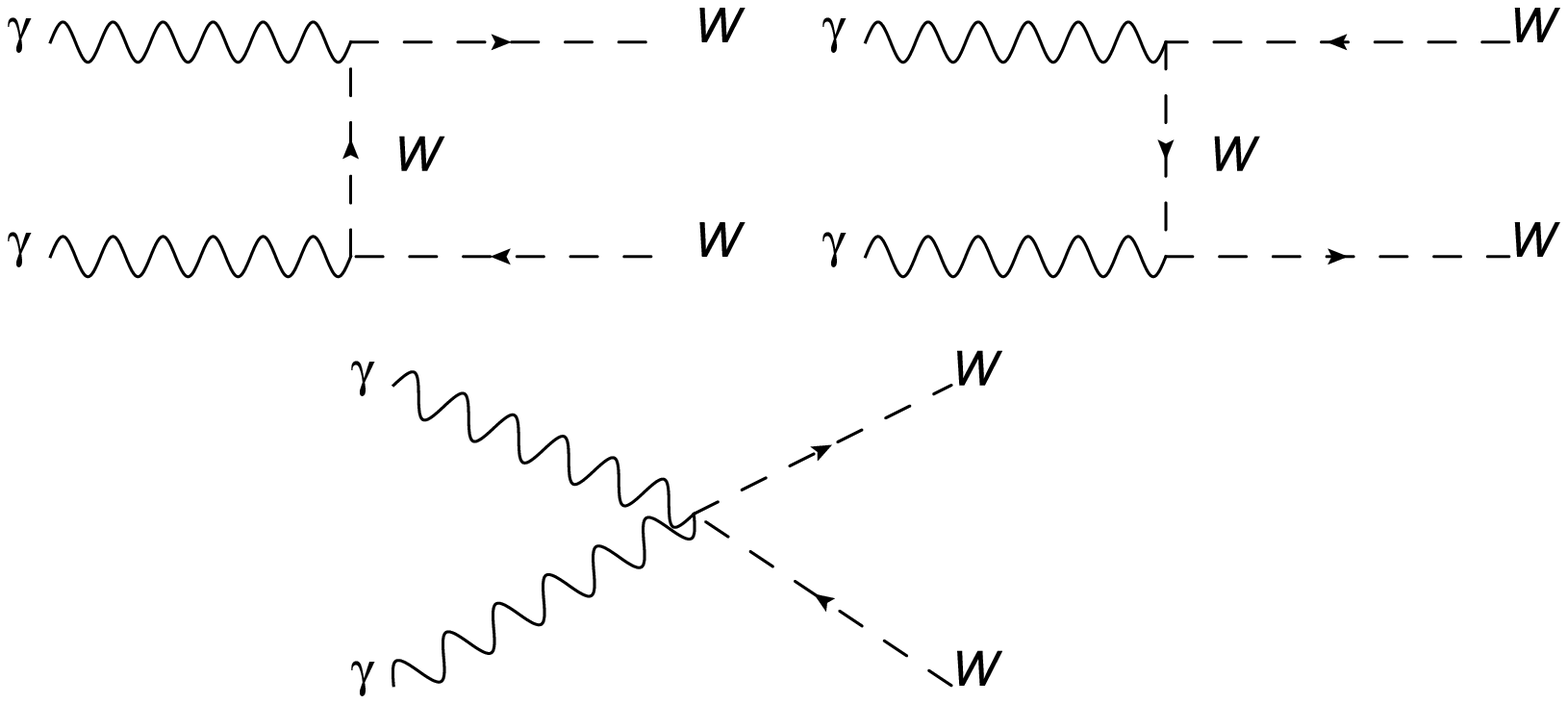}
\caption{Tree-level Feynman diagrams for the subprocess $\gamma^{*} \gamma^{*}\rightarrow W^{-} W^{+}$.
\label{fig5}}
\end{figure}

\begin{figure}
\includegraphics{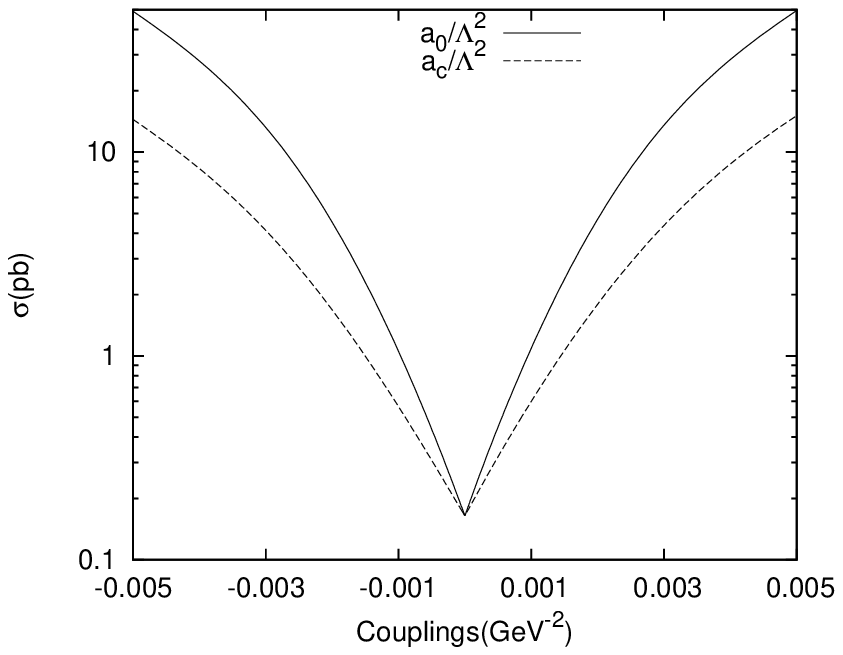}
\caption{The total cross sections as function of $\frac{a_{0}}{\Lambda^{2}}$ and $\frac{a_{c}}{\Lambda^{2}}$ aQGCs for the $e^{+}e^{-}\rightarrow W^{-} W^{+}\gamma$ at the CLIC with $\sqrt{s}=0.5$ TeV.
\label{fig6}}
\end{figure}

\begin{figure}
\includegraphics{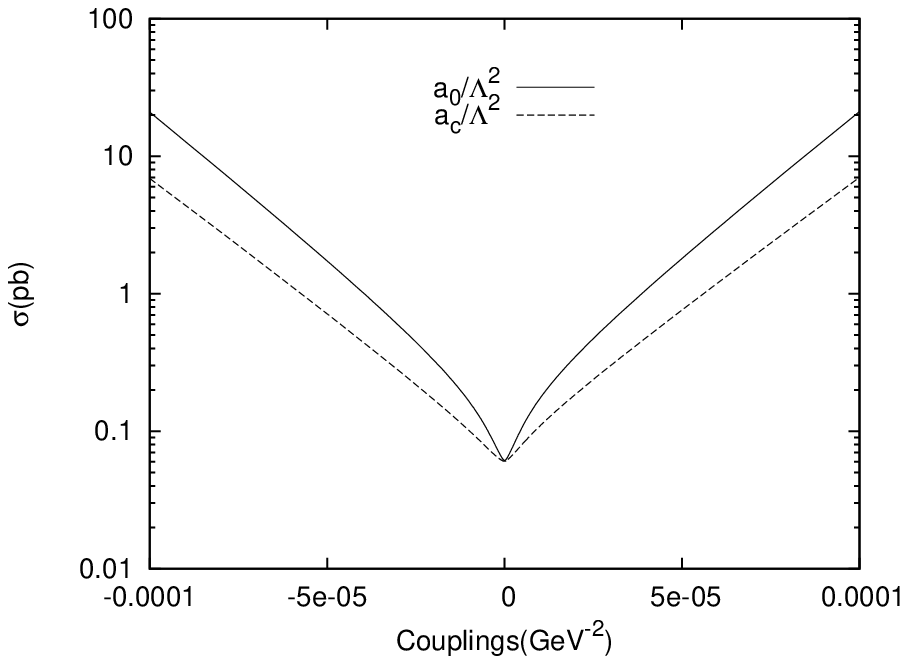}
\caption{The same as Fig. $6$ but for $\sqrt{s}=1.5$ TeV.
\label{fig7}}
\end{figure}

\begin{figure}
\includegraphics{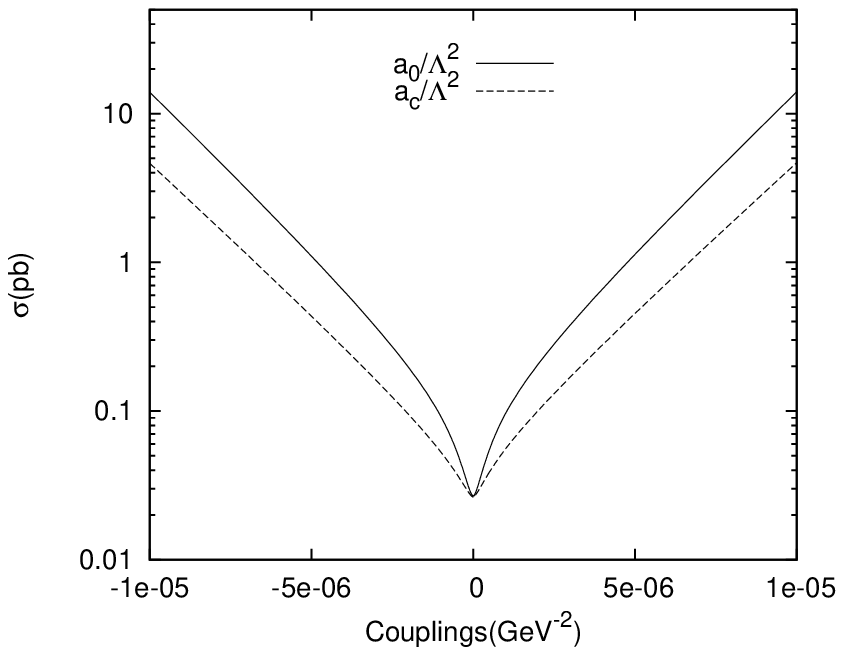}
\caption{The same as Fig. $6$ but for $\sqrt{s}=3$ TeV.
\label{fig8}}
\end{figure}

\begin{figure}
\includegraphics{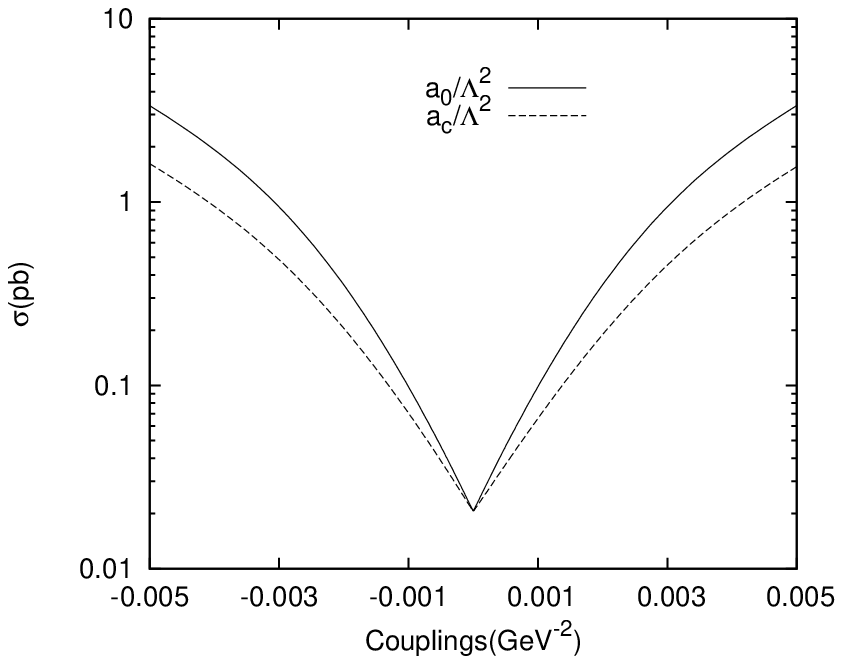}
\caption{The total cross sections as function of $\frac{a_{0}}{\Lambda^{2}}$ and $\frac{a_{c}}{\Lambda^{2}}$ aQGCs for the $e^{+}e^{-} \rightarrow e^{+}\gamma^{*} e^{-} \rightarrow e^{+} W^{-} \gamma \nu_{e}$ at the CLIC with $\sqrt{s}=0.5$ TeV.
\label{fig9}}
\end{figure}

\begin{figure}
\includegraphics{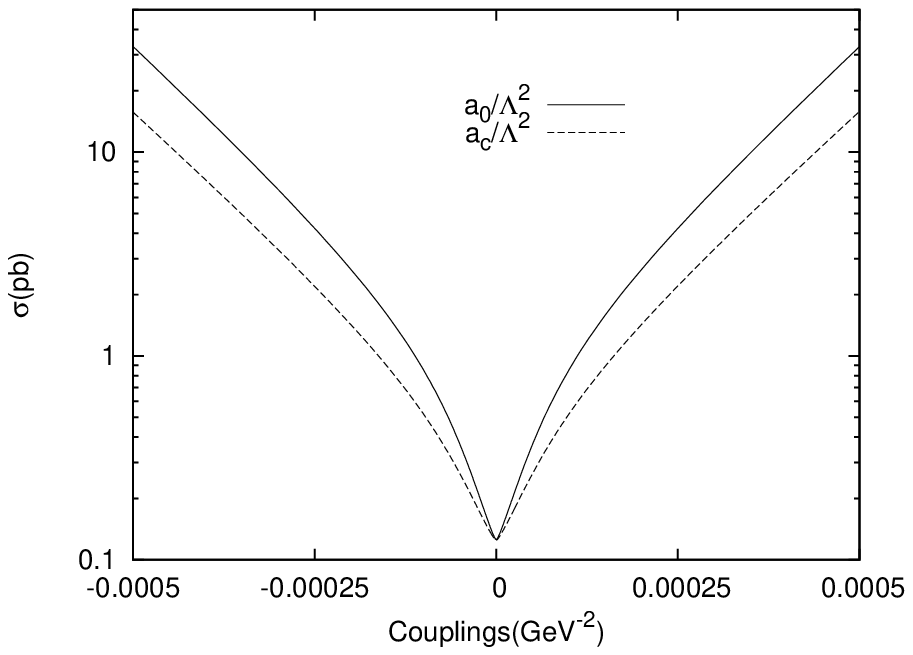}
\caption{The same as Fig. $9$ but for $\sqrt{s}=1.5$ TeV.
\label{fig10}}
\end{figure}

\begin{figure}
\includegraphics{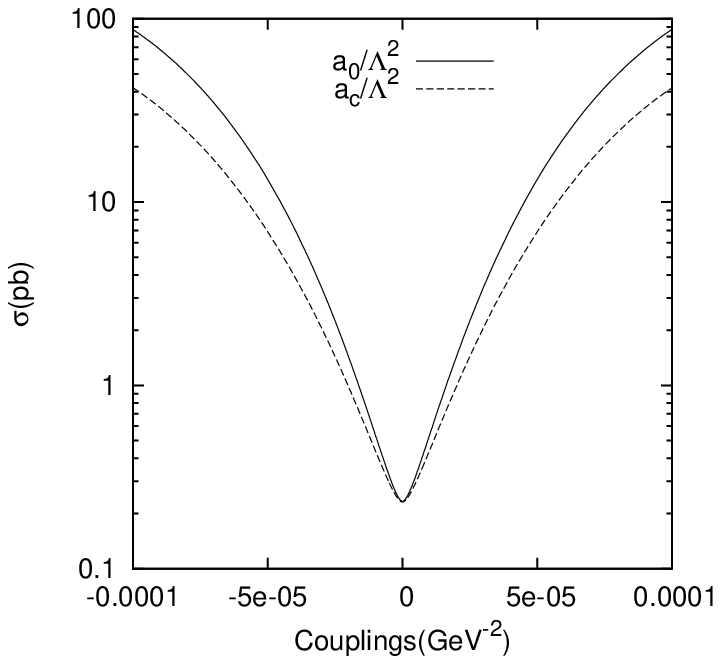}
\caption{The same as Fig. $9$ but for $\sqrt{s}=3$ TeV.
\label{fig11}}
\end{figure}

\begin{figure}
\includegraphics{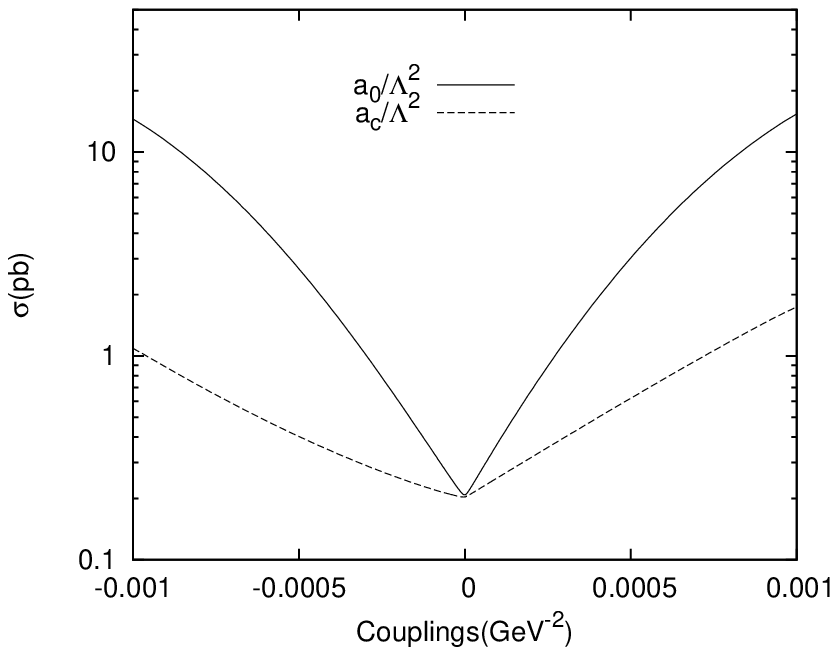}
\caption{The total cross sections as function of $\frac{a_{0}}{\Lambda^{2}}$ and $\frac{a_{c}}{\Lambda^{2}}$ aQGCs for the $e^{+}e^{-} \rightarrow e^{+}\gamma^{*} \gamma^{*} e^{-} \rightarrow e^{+} W^{-} W^{+} e^{-}$ at the CLIC with $\sqrt{s}=0.5$ TeV.
\label{fig12}}
\end{figure}

\begin{figure}
\includegraphics{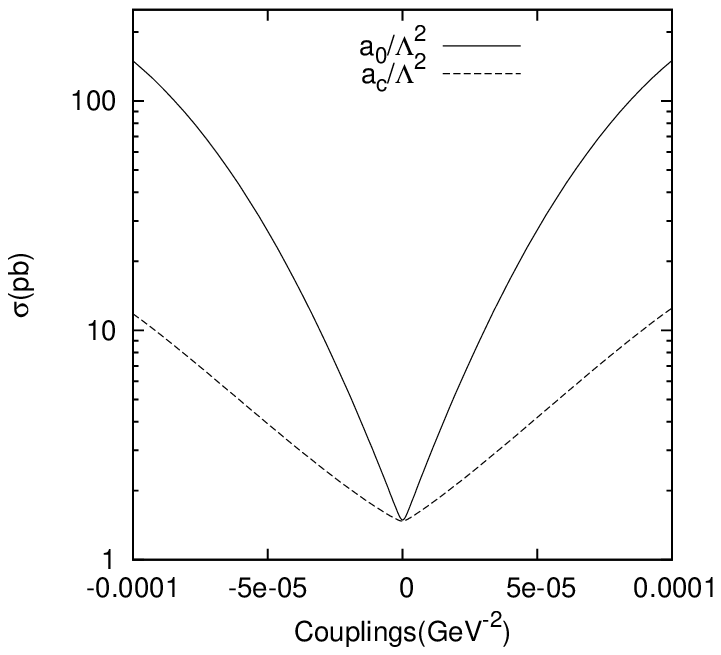}
\caption{The same as Fig. $12$ but for $\sqrt{s}=1.5$ TeV.
\label{fig13}}
\end{figure}

\begin{figure}
\includegraphics{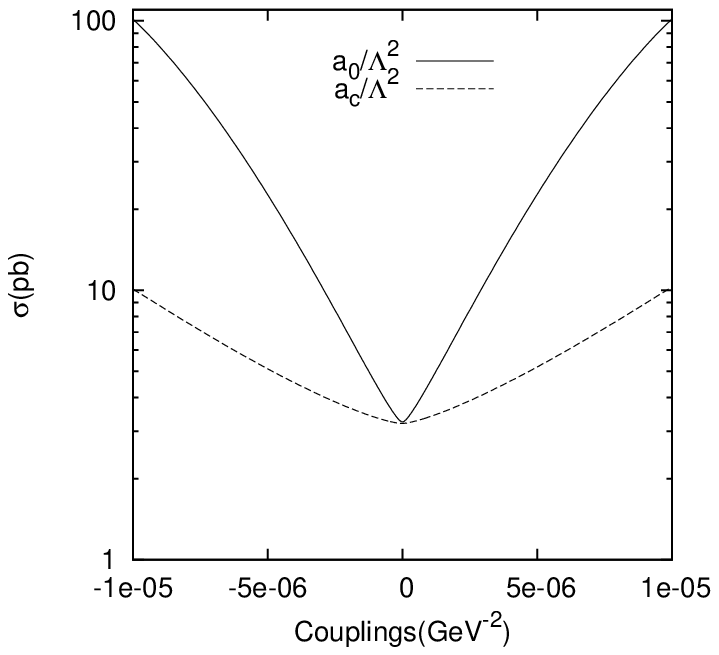}
\caption{The same as Fig. $12$ but for $\sqrt{s}=3$ TeV.
\label{fig14}}
\end{figure}

\begin{table}
\caption{The fundamental parameters of the three energy options of the CLIC. Here $\sqrt{s}$ is the center-of-mass energy, $N$ is the number of particles in bunch, $L$ is the total luminosity, $\sigma_{x,y,z}$ are the average sizes of the bunches \cite{18}.
\label{tab1}}
\begin{ruledtabular}
\begin{tabular}{ccccc}
Parameter& Unit& Stage\,$1$& Stage\,$2$& Stage\,$3$ \\
\hline
$\sqrt{s}$& TeV& $0.5$& $1.5$& $3$ \\
$N$& $10^{9}$& $3.7$& $3.7$& $3.7$  \\
$L$& fb$^{-1}$& $230$& $320$& $590$  \\
$\sigma_{x}$& nm& $100$& $60$& $40$ \\
$\sigma_{y}$& nm& $2.6$& $1.5$& $1$  \\
$\sigma_{z}$& $\mu$m& $44$& $44$& $44$  \\
\end{tabular}
\end{ruledtabular}
\end{table}

\begin{table}
\caption{$95\%$ C.L. sensitivity limits of the $\frac{a_{0}}{\Lambda^{2}}$ and $\frac{a_{c}}{\Lambda^{2}}$ aQGCs through the processes $e^{+}e^{-}\rightarrow W^{-} W^{+}\gamma$ at the CLIC with $\sqrt{s}=0.5,1.5$ and $3$ TeV.
\label{tab2}}
\begin{ruledtabular}
\begin{tabular}{cccc}
$\sqrt{s}$ (TeV)& $L_{int}$(fb$^{-1}$)& $\frac{a_{0}}{\Lambda^{2}}$(TeV$^{-2}$)& $\frac{a_{c}}{\Lambda^{2}}$ (TeV$^{-2}$)\\
\hline
$0.5$& $10$& $[-120;\, 93]$& $[-253;\, 141]$ \\
$0.5$& $50$& $[-86.5;\, 54.4]$& $[-195;\, 81.3]$  \\
$0.5$& $100$& $[-76.3;\, 44.1]$& $[-176;\, 65.0]$\\
$0.5$& $230$& $[-65.6;\, 33.7]$& $[-160;\, 49.3]$ \\
\hline
$1.5$& $10$& $[-2.66;\, 2.25]$& $[-4.69;\, 3.83]$ \\
$1.5$& $100$& $[-1.53;\, 1.16]$& $[-2.82;\, 1.98]$  \\
$1.5$& $200$& $[-1.37;\, 0.95]$& $[-2.47;\, 1.61]$\\
$1.5$& $320$& $[-1.23;\, 0.83]$& $[-2.25;\, 1.38]$ \\
\hline
$3$& $10$& $[-0.26;\, 0.23]$& $[-0.45;\, 0.40]$ \\
$3$& $100$& $[-0.15;\, 0.13]$& $[-0.26;\, 0.22]$  \\
$3$& $300$& $[-0.12;\, 0.09]$& $[-0.21;\, 0.16]$\\
$3$& $590$& $[-0.10;\, 0.07]$& $[-0.18;\, 0.13]$ \\
\end{tabular}
\end{ruledtabular}
\end{table}

\begin{table}
\caption{$95\%$ C.L. sensitivity limits of the $\frac{a_{0}}{\Lambda^{2}}$ and $\frac{a_{c}}{\Lambda^{2}}$ aQGCs through the processes
 $e^{+}e^{-} \rightarrow e^{+}\gamma^{*} e^{-} \rightarrow e^{+} W^{-} \gamma \nu_{e}$ at the  CLIC with $\sqrt{s}=0.5,1.5$ and $3$ TeV.
\label{tab3}}
\begin{ruledtabular}
\begin{tabular}{cccc}
$\sqrt{s}$ (TeV)& $L_{int}$(fb$^{-1}$)& $\frac{a_{0}}{\Lambda^{2}}$(TeV$^{-2}$)& $\frac{a_{c}}{\Lambda^{2}}$ (TeV$^{-2}$)\\
\hline
$0.5$& $10$& $[-162;\, 158]$& $[-191;\, 286]$ \\
$0.5$& $50$& $[-109;\, 104]$& $[-116;\, 211]$  \\
$0.5$& $100$& $[-89.0;\, 86.0]$& $[-93.0;\, 184]$\\
$0.5$& $230$& $[-73.0;\, 71.0]$& $[-68.0;\, 161]$ \\
\hline
$1.5$& $10$& $[-8.08;\, 8.02]$& $[-10.9;\, 12.7]$ \\
$1.5$& $50$& $[-4.52;\, 4.48]$& $[-5.41;\, 7.45]$  \\
$1.5$& $100$& $[-3.75;\, 3.71]$& $[-4.57;\, 6.63]$\\
$1.5$& $230$& $[-3.29;\, 3.24]$& $[-4.06;\, 6.12]$ \\
\hline
$3$& $10$& $[-1.15;\, 1.12]$& $[-1.59;\, 1.71]$ \\
$3$& $100$& $[-0.64;\, 0.63]$& $[-0.83;\, 0.98]$  \\
$3$& $300$& $[-0.50;\, 0.49]$& $[-0.61;\, 0.76]$\\
$3$& $590$& $[-0.41;\, 0.40]$& $[-0.48;\, 0.68]$ \\
\end{tabular}
\end{ruledtabular}
\end{table}

\begin{table}
\caption{$95\%$ C.L. sensitivity limits of the $\frac{a_{0}}{\Lambda^{2}}$ and $\frac{a_{c}}{\Lambda^{2}}$ aQGCs for $e^{+}e^{-} \rightarrow e^{+}\gamma^{*} \gamma^{*} e^{-} \rightarrow e^{+} W^{-} W^{+} e^{-}$ at the CLIC with $\sqrt{s}=0.5,1.5$ and $3$ TeV.
\label{tab4}}
\begin{ruledtabular}
\begin{tabular}{cccc}
$\sqrt{s}$ (TeV)& $L_{int}$(fb$^{-1}$)& $\frac{a_{0}}{\Lambda^{2}}$(GeV$^{-2}$)& $\frac{a_{c}}{\Lambda^{2}}$ (GeV$^{-2}$)\\
\hline
$0.5$& $10$& $[-57.8;\,27.8]$& $[-321;\,59.1]$ \\
$0.5$& $50$& $[-45.8;\,15.4]$& $[-293;\,29.5]$  \\
$0.5$& $100$& $[-42.5;\,11.8]$& $[-286;\,21.5]$\\
$0.5$& $230$& $[-38.9;\,8.50]$& $[-279;\,14.8]$ \\
\hline
$1.5$& $10$& $[-2.21;\,1.90]$& $[-9.38;\,6.22]$ \\
$1.5$& $100$& $[-1.31;\,1.03]$& $[-6.07;\,2.96]$  \\
$1.5$& $200$& $[-1.14;\,0.83]$& $[-5.57;\,2.42]$\\
$1.5$& $320$& $[-1.04;\,0.75]$& $[-5.18;\,1.95]$ \\
\hline
$3$& $10$& $[-0.31;\,0.30]$& $[-1.22;\,1.08]$ \\
$3$& $100$& $[-0.18;\,0.17]$& $[-0.73;\,0.59]$  \\
$3$& $300$& $[-0.14;\,0.13]$& $[-0.59;\,0.44]$\\
$3$& $590$& $[-0.11;\,0.10]$& $[-0.50;\,0.34]$ \\
\end{tabular}
\end{ruledtabular}
\end{table}

\end{document}